\documentclass[prd,twocolumn,showpacs,preprintnumbers,amsmath,amssymb,citeautoscript,nofootinbib]{revtex4}
\usepackage[dvips]{epsfig}
\usepackage[usenames]{color}

\newcommand{\nc}[1]{\newcommand{#1}}

\nc{\its}[1]{\itshape #1 \upshape}
\nc{\mc}[3]{\multicolumn{#1}{#2}{#3}}

\nc{\bc}{\begin{center}}
\nc{\ec}{\end{center}}
\nc{\ig}[1]{\bc \includegraphics{#1} \ec}

\nc{\bo}[1]{\mbox{\boldmath \( #1 \! \! \)  \unboldmath}}
\newcommand{\beqn} {\begin{equation}}
\newcommand{\eqn} {\end{equation}}
\nc{\be}{\begin{eqnarray}}
\nc{\ee}{\end{eqnarray}}
\nc{\bew}{\begin{eqnarray*}}
\nc{\eew}{\end{eqnarray*}}
\nc{\bs}{\begin{subeqnarray}}   
\nc{\es}{\end{subeqnarray}}     
\nc{\nnn}{\nonumber \\}
\nc{\f}[2]{\frac{#1}{#2}}
\nc{\td}[2]{\f{d #1}{d #2}}
\nc{\pd}[2]{\f{\partial #1}{\partial #2}}
\nc{\suli}{\sum\limits}
\nc{\proli}{\prod\limits}
\nc{\ili}{\int\limits}
\nc{\sr}[2]{\stackrel{#1}{#2}}
\nc{\dps}{\displaystyle}
\nc{\ket}[1]{\left| #1 \right>}
\nc{\bra}[1]{\left< #1 \right|}
\nc{\bracket}[2]{\left< #1 \right| \left. \! #2 \right>}
\nc{\norm}[1]{\left\| #1 \right\|}
\nc{\lndm}[1]{\pd{^{#1} \ln{\det{M}}}{\mu^{#1}}}
\nc{\pdmm}[1]{M^{-1} \pd{^{#1} M}{\mu^{#1}}}
\nc{\pdm}{M^{-1}\pd{M}{\mu}}
\nc{\trac}[1]{\mbox{Tr}\left(#1\right)}
\nc{\hm}{\hat{m}}
\nc{\hmu}{\hat{\mu}}

\def\lsim{\raise0.3ex\hbox{$<$\kern-0.75em\raise-1.1ex\hbox{$\sim$}}}
\def\gsim{\raise0.3ex\hbox{$>$\kern-0.75em\raise-1.1ex\hbox{$\sim$}}}

\begin{document}

\title{Equation of State for physical quark masses}

\author{M. Cheng$^{\rm a}$, S. Ejiri$^{\rm b}$, P. Hegde$^{\rm b,c}$, F. Karsch$^{\rm b,d}$, 
O. Kaczmarek$^{\rm d}$, E. Laermann$^{\rm d}$,\\
R. D. Mawhinney$^{\rm e}$, C. Miao$^{\rm b}$, S. Mukherjee$^{\rm b}$,
P. Petreczky$^{\rm b,f}$, C. Schmidt$^{\rm d}$, W. Soeldner$^{\rm g}$ 
}

\affiliation{
$^{\rm a}$ Lawrence Livermore National Laboratory,
Livermore, CA 94550, USA\\
$^{\rm b}$ Physics Department, Brookhaven National Laboratory,
Upton, NY 11973, USA \\
$^{\rm c}$ Department of Physics and Astronomy, Stony Brook
University, Stony Brook, NY 11790, USA\\
$^{\rm d}$ Fakult\"at f\"ur Physik, Universit\"at Bielefeld, D-33615 Bielefeld,
Germany\\
$^{\rm e}$ Physics Department,Columbia University, New York, NY 10027, USA\\
$^{\rm f}$ RIKEN-BNL Research Center, Brookhaven National Laboratory,
Upton, NY 11973, USA \\
$^{\rm g}$  ExtreMe Matter Institute EMMI, GSI Helmholtzzentrum f\"ur Schwerionenforschung,
Planckstrasse~1, D-64291 Darmstadt, Germany
}

\date{\today}
\preprint{BI-TP 2009/28}
\preprint{CU-TP-1190}

\begin{abstract}
We calculate the QCD equation of state for temperatures corresponding 
to the transition region with physical mass values for two degenerate light
quark flavors and a strange quark using an improved staggered fermion
action (p4-action) on lattices with temporal extent $N_{\tau}=8$.
We compare our results with previous calculations performed at twice
larger values of the light quark masses as well as with results obtained
from a resonance gas model calculation. We also discuss the deconfining and
chiral aspects of the QCD transition in terms of renormalized Polyakov loop,
strangeness fluctuations and subtracted chiral condensate. We show that compared
to the calculations performed at twice larger value of the light quark mass
the transition region shifts by about $5$MeV toward smaller temperatures.
\end{abstract}


\maketitle

\section{Introduction}
\label{intro}
First calculations of the equation of state (EoS) of hot strongly interacting matter 
date back to the early 80s \cite{su2,su3}. These early purely gluonic calculations have 
been improved steadily by including contributions from dynamical quark degrees of freedom 
(for recent reviews see \cite{detar_lat08,petreczky_sewm06,petreczky_qm09}). 
In the most recent calculations the equation of state has been evaluated
for 2+1 flavor QCD, {\it i. e.} in QCD with one strange quark and two light ($u,d$) quarks using
various improved staggered fermion actions \cite{stout,MILCeos,ourEoS,hotqcd_eos}. 
The most extensive
calculations of the EoS have been performed with p4 and asqtad staggered fermion
formulations on
lattices with temporal extent $N_{\tau}=4,~6$ \cite{MILCeos,ourEoS} and $8$ \cite{hotqcd_eos}.
These actions improve both the flavor symmetry of the staggered fermions as well as the 
quark dispersion relations. The latter insures that thermodynamic observables are
${\cal O}(a^2)$ improved at high temperatures and thus have only a small cut-off 
dependence in this regime. The stout-link action, which has been used for the calculation
of the EoS on lattices with temporal extent $N_{\tau}=4,~6$ \cite{stout}, only
improves the flavor symmetry of the staggered fermions and therefore has the same large
discretization errors at high temperatures as the standard staggered fermion formulation. 
Indeed, the calculations of the EoS with the stout-link action on lattices with 
temporal extent $N_{\tau}=4$ and $6$ reflect the expected strong cutoff dependence above 
the  transition temperature \cite{stout}.

While at high temperatures the masses of the relevant degrees of freedom, quarks and gluons,
are small compared to the temperature scale, this is not the case at low temperatures
and in the transition region. One thus may expect that at these temperatures thermodynamic 
observables are more sensitive to the quark masses, which control the mass of the
light pseudo-scalars and eventually are responsible for the occurrence of a true
phase transition in the chiral limit. Calculations with p4 and asqtad actions have so-far
been performed using light quark masses ($\hm_l$) which are one tenth of the strange 
quark mass ($\hm_s$) and correspond to a pseudo-scalar Goldstone 
mass\footnote{In calculations with staggered fermions 
flavor symmetry is broken at non-vanishing values of the lattice spacing $a$. As a 
consequence only one of the pseudo-scalar mesons has a light mass that is
proportional to $\sqrt{m_l}$ and vanishes in the chiral limit at fixed
$a >0$. Full chiral symmetry with the correct Goldstone pion multiplet
is recovered only for $a\rightarrow 0$.
For an estimate of the remaining flavor symmetry violations in 
spectrum calculations with the p4-action see~\cite{FlavBreak}.}
of $220$ and $260$ MeV respectively \cite{hotqcd_eos}.
The calculations with the stout-link action have been performed at the physical value of 
the light quark mass.

The purpose of this paper is to investigate the quark mass dependence of the EoS by 
calculating it with the p4-action for physical values of the (degenerate) light quark 
masses. The calculational procedure used in this work
closely follows that used in our previous calculations at $\hm_l=0.1\hm_s$ \cite{ourEoS}. 
The 
paper is organized as follows.
In the next section we discuss the parameters and some technical details of the numerical calculations,
including the choice of the quark masses and the determination of the lattice 
spacing that fixes the temperature scale, $aT=1/N_\tau$. In section III we 
show the QCD equation of state and compare it with the resonance gas model.
In section IV we discuss chiral and deconfining aspects of the QCD transition. 
Finally section V contains our conclusions.
In the appendix we give further numerical details of our calculations.


\section{Zero temperature calculations}
We have performed calculations with the p4-action for several values of the gauge coupling
$\beta=6/g^2$ in the region of the finite temperature crossover for lattices with
temporal extent $N_{\tau}=8$.
The values of the quark masses
and gauge coupling are given in Table \ref{tab:t=0}. Simulations were performed with the
Rational Hybrid Monte-Carlo (RHMC) algorithm with Hasenbush preconditioning and different time
steps in the molecular dynamic (MD) evolution for gauge fields, strange quarks and light quarks.
We used 5 time steps for strange quark updates per each light quark update. The gauge fields
were updated 10 times more frequently than the strange quarks. The length of the 
MD trajectory was $0.5$ and the acceptance rate about $70\%$. 
The zero temperature calculations were performed on $32^4$ lattices. 

We would like to perform calculations of thermodynamic quantities keeping the 
physical values of the strange quark mass
as well as the light quark masses fixed.
As a starting point for our calculations we used 
the same fine-tuned values of the bare strange quark mass as in \cite{ourEoS}
but bare light quark masses
which are twice smaller than in Ref. \cite{ourEoS},
i.e. the ratio of strange to light quark mass was chosen to be $h=m_s/m_l=20$.

At each beta value we then have determined 
the static quark potential.
The static potential $V_{q \bar q}(r)$ and the corresponding scales $r_0$ and $r_1$ defined
as 
\begin{equation}
\left( r^2 \frac{d V_{q\bar q}(r)}{d r} \right)_{r=r_n}=
\begin{cases}
1.65\;,& n=0\\
1.0~\;,& n=1
\end{cases}
\end{equation}
were calculated from smeared Wilson loops as described in Ref. \cite{ourTc}. Here
we used 30 levels of APE smearing of the spatial links to get good signals for the Wilson loop 
expectation values. The values
of $r_0/a$ and $r_1$/a are given in Table \ref{tab:t=0}. 
The comparison with previous results given in Table I of \cite{ourEoS}
reveals that the new values
of the scale parameters in lattice units differ from the old ones
obtained at $h=10$ by less than 1 \%, with deviations which are not systematic.
Thus, the smaller values of the light quark masses do not change the lattice spacing
in physical units beyond the errors, 
which are about $0.5\%$ (c.f. Table \ref{tab:t=0}). As the consequence the values of the 
temperature  in units of $r_0$ have errors of about $0.5\%$. To give the temperature
in physical units, e.g. MeV, we use the value $r_0=0.469(7)$fm. Thus the temperature
scale in physical units has an overall uncertainty of about $1.5\%$. However, since the
same uncertainty is present in the calculations performed at twice larger quark mass it 
does not show up in the comparison of thermodynamic quantities obtained at two different quark masses.

To remove the additive divergent constant $c_0$ in the potential, following Ref.
\cite{ourEoS} we normalized it to the string form of the static quark
potential, $V_{\rm string} (r) = -\pi/12 r + \sigma r$, at distance $r=1.5r_0$.
This amounts to renormalize the temporal gauge links
with $z(\beta) = \exp(c_0/2)$. The resulting multiplicative renormalization
factors, $z(\beta)$, are also given in Table~\ref{tab:t=0}
and will be used to define renormalized Polyakov loops that are discussed
in section IV.

Having determined the scale 
we extracted pseudo-scalar meson masses 
using wall sources in the calculation of meson propagators.
It turned out that the $\eta_{s \bar s}$ mass in lattice units
is, with 1 -2 \% accuracy, the same as in \cite{ourEoS}.
Thus, 
a re-adjustment of the line of constant physical $\eta_{s \bar s}$ mass 
corresponding to our new and smaller light quark masses was not
necessary. 
In fact,
in the present calculations we find that our quark mass values
define a line of constant physics characterized by
the following relations
\begin{eqnarray}
r_0 \cdot m_{\pi}=0.371(3)&,& r_0 \cdot m_K=1.158(5),\nonumber \\
r_0 \cdot m_{\eta_{s\bar s}}&=&1.578(7)\; .
\label{lcp}
\end{eqnarray}
Using $r_0=0.469$ fm, as determined in Ref.\cite{gray}, we get $m_{\pi}=154$ MeV, 
$m_K=486$ MeV and\footnote{ A physical value for
the $\eta_{s \bar s}$ mass can be obtained from the relation 
$m_{\eta_{s\bar s}}=\sqrt{2 m_K^2-m_{\pi}^2} = 686$ MeV.}
$m_{\eta_{s\bar s}}=663$~MeV. 
This means that both the light quark masses
and the strange quark mass are very close to their physical values. Furthermore, 
in the entire parameter range covered by our thermodynamic calculations deviations of the 
meson masses from the above values are less than $3\%$.

\begin{table*}
\begin{tabular}{|c|c|c|c|c|c|c|c|c|c|}
\hline
$\beta$ & $\hat m_s$ &  $\# traj.$  & $r_0/a$   &  $r_1/a$   & $m_{\pi}a$& $m_Ka$  & $m_{\eta_{s \bar s}} a$ & 
$z(\beta)$ \\
\hline
3.430  & 0.0370     &  2610 & 2.6698(101) & 1.8378(184) & 0.1415(2) & 0.4423(3) & 0.5992(4)  & 1.5124(90) \\
3.460  & 0.0313     &  3220 & 2.9377(90)  & 2.0000(202) & 0.1274(4) & 0.3994(4) & 0.5427(2)  & 1.5219(71) \\
3.490  & 0.0290     &  3460 & 3.2188(70)  & 2.2115(101) & 0.1213(15)& 0.3775(5) & 0.5117(10) & 1.5277(26) \\
3.500  & 0.0253     &  3030 & 3.3251(114) & 2.2550(159) & 0.1110(5) & 0.3491(4) & 0.4743(3)  & 1.5292(50) \\
3.510  & 0.0260     &  3040 & 3.4021(141) & 2.3116(105) & 0.1121(10)& 0.3510(9) & 0.4770(6)  & 1.5289(55) \\
3.520  & 0.0240     &  2980 & 3.5167(68)  & 2.4050(106) & 0.1056(5) & 0.3322(6) & 0.4525(3)  & 1.5303(20) \\
3.530  & 0.0240     &  2450 & 3.5927(101) & 2.4723(81)  & 0.1045(14)& 0.3300(17)& 0.4500(11) & 1.5291(48) \\
3.540  & 0.0240     &  3360 & 3.6802(138) & 2.5090(94)  & 0.1040(4) & 0.3251(13)& 0.4432(11) & 1.5272(41) \\
3.545  & 0.0215     &  3090 & 3.7513(67)  & 2.5840(75)  & 0.0947(7) & 0.3061(3) & 0.4178(2)  & 1.5298(23) \\ 
3.560  & 0.0205     &  3010 & 3.8790(67)  & 2.6732(81)  & 0.0935(12)& 0.2941(13)& 0.4024(11) & 1.5275(18) \\
3.585  & 0.0192     &  1400 & 4.1501(118) & 2.8503(106) & 0.1049(6) & 0.3258(17)& 0.4434(11) & 1.5275(66) \\
3.600  & 0.0192     &  4080 & 4.3033(181) & 2.9351(73)  & 0.0888(20)& 0.2731(6) & 0.3722(7)  & 1.5277(41) \\
3.630  & 0.0170     &  2000 & 4.6853(349) & 3.1858(167) & 0.0773(24)& 0.2442(5) & 0.3343(9)  & 1.5293(82) \\
3.660  & 0.0170     &  2850 & 4.8497(232) & 3.3368(138) & 0.0757(7) & 0.2364(12)& 0.3246(8)  & 1.5175(50) \\
\hline
\end{tabular}
\caption{The parameters of the zero temperature calculations, the values of $r_0$ and $r_1$ 
and the pseudo-scalar meson masses. The last column shows the normalization constant for the static potential.
}
\label{tab:t=0}
\end{table*}

\section{Calculation of the thermodynamic quantities}
The calculation of the EoS starts with the evaluation of
the trace anomaly, {\it i.e.} the trace of the energy-momentum tensor $\Theta_{\mu\mu}(T)$.
It is related to the temperature derivative of the pressure through thermodynamic identities, 
\begin{equation}
\frac{\Theta_{\mu\mu}(T)}{T^4}=\frac{\epsilon-3p}{T^4}=T\frac{{\rm d}}{{\rm d}T} \left( \frac{p}{T^4}\right)\; .
\label{e-3p}
\end{equation}
The trace anomaly can be expressed in terms of the expectation values of quark condensates
and the gluon action density
\begin{eqnarray}
&
\displaystyle
\frac{\Theta_{\mu\mu}(T)}{T^4}=\frac{\Theta^{\mu\mu}_G(T)}{T^4} +
\frac{\Theta^{\mu\mu}_F(T)}{T^4}\\[2mm]
&
\displaystyle
\frac{\Theta^{\mu\mu}_G(T)}{T^4} =
R_\beta
\left[ \langle s_G \rangle_0 - \langle s_G \rangle_\tau \right] N_\tau^4 \; , 
\label{e3pgluon}  \\[2mm]
&
\displaystyle
\frac{\Theta^{\mu\mu}_F(T)}{T^4} = - R_\beta R_{m}
\left[ 2 \hm_l
\left(\langle\bar{\psi}\psi \rangle_{l,0}
- \langle\bar{\psi}\psi \rangle_{l,\tau}\right) \right . \nonumber\\[2mm]
&
\displaystyle
\left .
+ \hm_s \left(\langle\bar{\psi}\psi \rangle_{s,0}
- \langle\bar{\psi}\psi \rangle_{s,\tau}\right)
\right] N_\tau^4 \; . \label{e3pfermion}
\end{eqnarray}
where the zero temperature expectation values are subtracted to render
the trace anomaly UV finite.
The expectation values of the light and strange quark condensates and the
action are defined as
\begin{eqnarray}
&
\displaystyle
\langle\bar{\psi}\psi\rangle_{q,x} \equiv \frac{1}{4}
\frac{1}{N_\sigma^3N_x} \left\langle {\rm Tr} D^{-1}(\hat{m}_q) 
\right\rangle_x \;\; ,\;\; \nonumber\\
&
\displaystyle
q=l,~s\;\; , \;\; x=0,~\tau \;\; ,
\label{quarkcondensate}\\
&
\displaystyle
\langle s_G\rangle_x \equiv \frac{1}{N_\sigma^3N_x} \left\langle S_G
\right\rangle_x \;\; ,
\label{gluondensity}
\end{eqnarray}
with $D(\hat{m}_q)$ being the staggered fermion matrix.
Furthermore, $R_\beta$ and $R_m$ are the non-perturbative beta function and the mass
anomalous dimension
\begin{eqnarray}
&
\displaystyle
R_{\beta}(\beta)=-a \frac{d \beta}{da}\\[2mm]
&
\displaystyle
R_m(\beta) = \frac{1}{\hm_l(\beta)}
\frac{{\rm d} \hm_l(\beta)}{{\rm d}\beta}.
\end{eqnarray} 
Following Ref. \cite{ourEoS} the $\beta$-dependence of the Sommer scale is fitted to a 
renormalization group inspired Ansatz \cite{allton}
\begin{eqnarray}
\frac{r_0}{a} = 
\displaystyle{\frac{1+ e_r \hat{a}^2 (\beta) +f_r \hat{a}^4 (\beta)}{a_r 
R_2(\beta) 
\left(1+ b_r \hat{a}^2 (\beta) +c_r \hat{a}^4 (\beta) +d_r \hat{a}^6 (\beta)
\right)}}
\;\;  
\label{fit}
\end{eqnarray}
where
\begin{eqnarray}
R_2(\beta)&=& \exp{\left(-\frac{\beta}{12b_0}\right)}
\left(\frac{6 b_0}{\beta}\right)^{-b_1/(2b_0^2)}
\end{eqnarray}
is the two-loop beta function for 3-flavor QCD and $\hat{a}(\beta) =  R_2(\beta)/R_2(3.4)$.
From this the non-perturbative beta function $R_{\beta}$ can be calculated as
\beqn
R_\beta (\beta) =  \frac{r_0}{a} \left(
\frac{{\rm d} r_0/a}{{\rm d}\beta} \right)^{-1}
\;\; .
\label{r0beta}
\eqn 
Likewise, $R_m$ was obtained from a parametrization of the bare quark mass,
\beqn
\hat m_l r_0/a = m^{RGI} r_0 \left(\frac{12 b_0}{\beta}\right)^{4/9} P(\beta)
\eqn
with a sixth order rational function $P(\beta)$ as in \cite{ourEoS}
to account for deviations
from the leading order scaling relation.

\begin{figure}
\includegraphics[width=10cm]{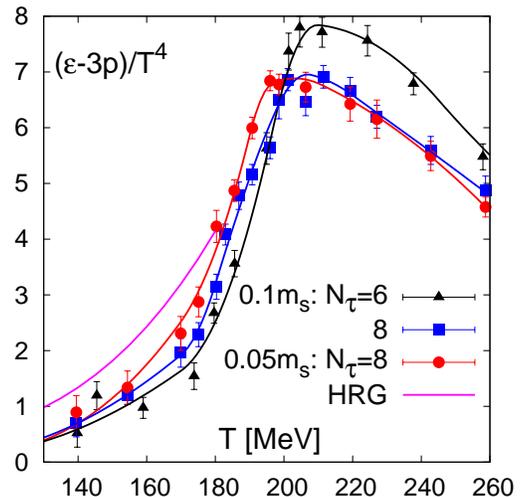}
\caption{The trace anomaly $(\epsilon-3p)/T^4$ calculated for
the physical quark mass and compared with previous calculations at larger
light quark masses $m_l=0.1m_s$ as well as with the HRG model which
includes all the resonances up to $2.5$GeV. Also shown are the interpolations
of the lattice data.}
\label{fig:e-3p}
\end{figure}

We performed finite temperature calculations on $32^3 \times 8$ lattices
for the 14 parameter sets shown in Table \ref{tab:t=0}. The number of MD trajectories
for each finite temperature run is given in Table \ref{tab:bare}.
Using the expectation values
of the quark condensates and the gluonic action density as well as the non-perturbative
beta functions described above we have calculated the trace anomaly. The numerical results
are shown in Figure \ref{fig:e-3p} and are compared to the previous calculation at twice
larger quark mass $m_l=0.1m_s$ on $N_{\tau}=6$ lattices \cite{ourEoS} and $N_{\tau}=8$
lattices \cite{hotqcd_eos}. The differences between $N_{\tau}=6$ and $N_{\tau}=8$ calculations
are due to cutoff effects and have been discussed in Ref. \cite{hotqcd_eos}.

As one can see from the figure 
the main differences to the $N_\tau=8$ results at $m_l = 0.1 m_s$
arise for temperatures $T \lsim 200$ MeV.
This difference can be understood as resulting from a shift of the transition
temperature when the light quark mass is lowered to approximately its physical
value. Based on our previous study on coarser lattices we expect that
the reduction of the quark mass from $0.1m_s$ to $0.05m_s$ effectively
leads to a shift of several observables by a few MeV towards smaller
values of the temperature \cite{ourTc}. As discussed further in section IV
this shift indeed amounts to about $5$~MeV.
At lower temperatures
it also is expected that the trace anomaly increases
with decreasing quark masses as hadrons become lighter when the quark mass is decreased.
While a tendency for such an increase may be indicated by the data at the lowest two
temperatures reached in our calculation, this effect is 
certainly not significant within
the current statistical accuracy.
The observed consistency within errors between the trace anomaly 
calculated at $m_l = 0.05 m_s$ and at $m_l = 0.1 m_s$
might possibly also hint towards a distortion of the hadron spectrum
due to finite lattice spacing effects.
In fact, it is known that for improved staggered fermions the ground state 
hadron masses approach their continuum limit from above \cite{milc01,milc04}.
For temperatures $T>200$ MeV there is no visible dependence of $(\epsilon-3p)/T^4$
on the quark mass. This is 
presumably due to the fact that hadronic degrees of freedom are no longer relevant in 
this temperature domain; the relevant degrees of freedom have thermal masses
of the order of the temperature and are insensitive to the light quark masses
already for $m_l=0.1m_s$.   

At temperatures below the transition temperature it is expected that thermodynamic 
quantities are well described by a hadron resonance gas (HRG) model. In fact, the freeze-out of 
hadrons in heavy ion experiments takes place in the transition region and the observed 
particle abundances are well described by the HRG model \cite{cleymans,pbm}. 
Therefore in Figure \ref{fig:e-3p} we also show the prediction of the HRG model, which
includes all the known resonances up to the mass $M_{max}=2.5$ GeV. 
The lattice data for $\epsilon-3p$ are below
the HRG prediction although the deviations from it are smaller compared to the results obtained
at $m_l=0.1m_s$. 
We mention again the present statistical accuracy and the possibility of
discretization effects in the hadron spectrum. In particular, due to taste breaking of staggered fermions
pseudo-scalar mesons are not degenerate at finite lattice spacing, therefore their contribution to thermodynamic
quantities maybe suppressed.

From the trace anomaly the pressure and thus other thermodynamic quantities can be calculated 
by performing the integration over the temperature
\beqn
\frac{p(T)}{T^4}-\frac{p(T_0)}{T_0^4}=\int_{T_0}^{T} d T' \frac{1}{{T'}^5} \Theta_{\mu \mu}(T').
\label{p_int}
\eqn
Here $T_0$ is an arbitrary temperature value that is usually chosen in the 
low temperature regime where the pressure and other thermodynamical quantities 
are suppressed exponentially by Boltzmann factors associated with the lightest
hadronic states, i.e. the pions.
Energy $\epsilon$ and entropy ($s T = (p+ \epsilon)$) 
densities are then obtained by combining results for $p/T^4$ and $(\epsilon-3p)/T^4$.

\begin{figure}
\includegraphics[width=8.8cm]{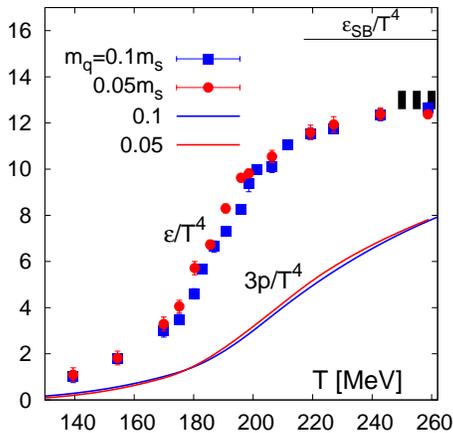}
\caption{Energy density and three times the pressure 
at the physical
value of the light quark mass and compared with 
previous calculations performed
at $m_l=0.1m_s$. The horizontal band shows the expected uncertainty in the energy
density due to the choice of the lower integration limit (see text). }
\label{fig:e+3p}
\end{figure}

To perform the integration numerically a reliable interpolation of the lattice data
on the trace anomaly is needed. These interpolations are shown in Figure~\ref{fig:e-3p}
as curves.
In the region $T \le 175$ MeV, the curves correspond to exponential fits.
Therefore our parametrization ensures that the pressure is zero at $T=T_0=0$.
However, the fits give very small values of the trace anomaly already at temperatures around $100$MeV, therefore
we could also use $T_0=100$MeV as lower integration limit.
Above $T = 175$ MeV, we divide the data into several intervals and perform quadratic interpolations. 
In each interval,
these quadratic fits have been adjusted to match the value and slope at the boundary with 
the previous interval. These interpolating curves are then used to calculate the pressure 
and other thermodynamic quantities using Eqs. (\ref{e-3p})
and (\ref{p_int}). 
The numerical results for the pressure and energy density are shown in Figure \ref{fig:e+3p}.
At temperatures between 170 MeV and 200 MeV,
energy density and pressure are slightly larger than
in earlier calculations performed at $m_l=0.1m_s$ \cite{hotqcd_eos}. 
For temperatures below $T<170$ MeV
the pressure is almost the same.  

Because the discretization errors increase as the temperature decreases
it is interesting to consider other choices for the normalization point $T_0$. 
At sufficiently low temperatures the HRG model provides a fair estimate for the pressure.
In particular, at $T=100$ MeV the pressure is not very sensitive on how many resonances
above $1.5$ GeV are included in the HRG model. Therefore we also calculated the pressure taking
the lower integration limit to be $T_0=100$ MeV with $p(T_0)/T_0^4=0.265$ calculated in the HRG model. 
The difference
in the pressure and energy density calculated this way and using the standard procedure, where $p(T_0)=0$,
can be used as an estimate of the systematic error. It is shown in Figure \ref{fig:e+3p} as the horizontal 
band in upper right corner. The thickness of the band indicates the expected size of the systematic error.
The systematic error estimated this way is about $8\%$
in the energy density
at the highest temperature of $T \simeq 260$MeV, and about $13\%$ for the pressure.

\section{Deconfinement and chiral aspects of the QCD transition}
\label{sec:deconf}

In this section we are going to discuss the deconfinement and chiral
aspects of the QCD transition in terms of renormalized Polyakov loop,
subtracted chiral condensate and strangeness susceptibility. The QCD transition
in terms of these quantities has been studied previously in 
Refs. \cite{ourEoS,hotqcd_eos,fodorTc}.

In the previous section we have seen that the energy density shows a rapid
rise in the temperature interval $T=(170-200)$MeV. This is
usually interpreted to be due to deconfinement, i.e.
liberation of many new degrees of freedom. For sufficiently large quark mass
this transition is known to be a first order transition (see e.g. Ref. \cite{stickan}).
In the limit of infinitely large quark mass the
order parameter for the deconfinement
phase transition  is the Polyakov loop. After renormalization it can be related
to the free energy of a static quark anti-quark pair $F_{\infty}(T)$ at infinite separation 
\cite{mclerran81,okacz02,plqcd}
\begin{equation}
L_{ren}(T)=\exp(-F_{\infty}(T)/(2 T)).
\end{equation}
A rapid change in this quantity  
is indicative for deconfinement
also in the presence of light quarks. 
The renormalized Polyakov loop is calculated from the bare Polyakov loop 
by multiplying
it by the renormalization constant $z(\beta)$ given in Table \ref{tab:t=0},
\begin{eqnarray}
&
L_{ren}(T)=z(\beta)^{N_{\tau}} L_{bare}(\beta)=\nonumber\\
&
z(\beta)^{N_{\tau}} \left<\frac{1}{3}  {\rm tr } 
\prod_{x_0=0}^{N_{\tau}-1} U_0(x_0,\vec{x})\right >
\end{eqnarray}
In the above formula $U_0(x_0,{\bf x})$ denotes the temporal link variables. Note that after
performing the ensemble average $L_{bare}(\beta)$ is independent of the space coordinate $\vec{x}$.

In the opposite limit of zero quark mass 
one expects a chiral transition
and the corresponding 
order parameter is the quark condensate defined in section III.
For a genuine phase transition, i.e. in the chiral limit
the quark condensate vanishes at the critical temperature $T_c$. However, we expect
that even for the crossover at finite quark mass the light quark
condensate rapidly drops in the transition region, indicating an
approximate restoration of the chiral symmetry. At non-vanishing quark mass
the quark condensate needs additive and multiplicative
renormalization. Therefore, following Ref. \cite{ourEoS} we introduce the so-called subtracted
chiral condensate
\begin{equation}
\Delta_{l,s}(T)=\frac{\langle \bar\psi \psi \rangle_{l,\tau}-\frac{m_l}{m_s} \langle \bar \psi \psi \rangle_{s,\tau}}
{\langle \bar \psi \psi \rangle_{l,0}-\frac{m_l}{m_s} \langle \bar \psi \psi \rangle_{s,0}}.
\end{equation}
Subtraction of the strange quark condensate multiplied by the ratio of the light to strange quark
mass removes the quadratic divergence proportional to the quark mass. 

\begin{figure}
\includegraphics[width=7cm]{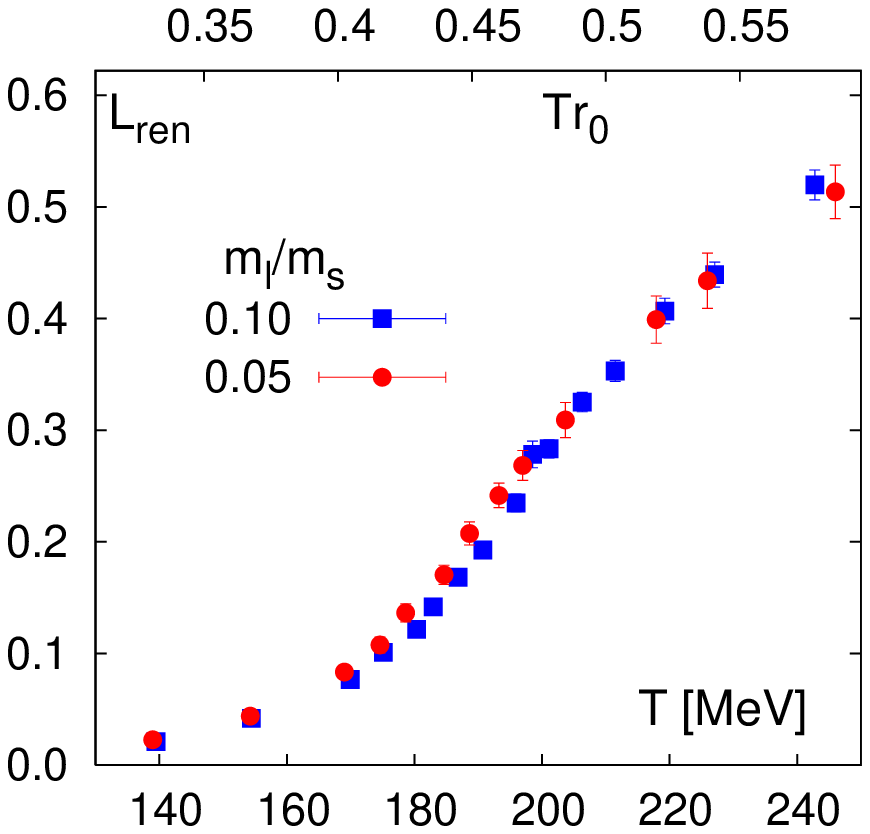}
\includegraphics[width=9.2cm]{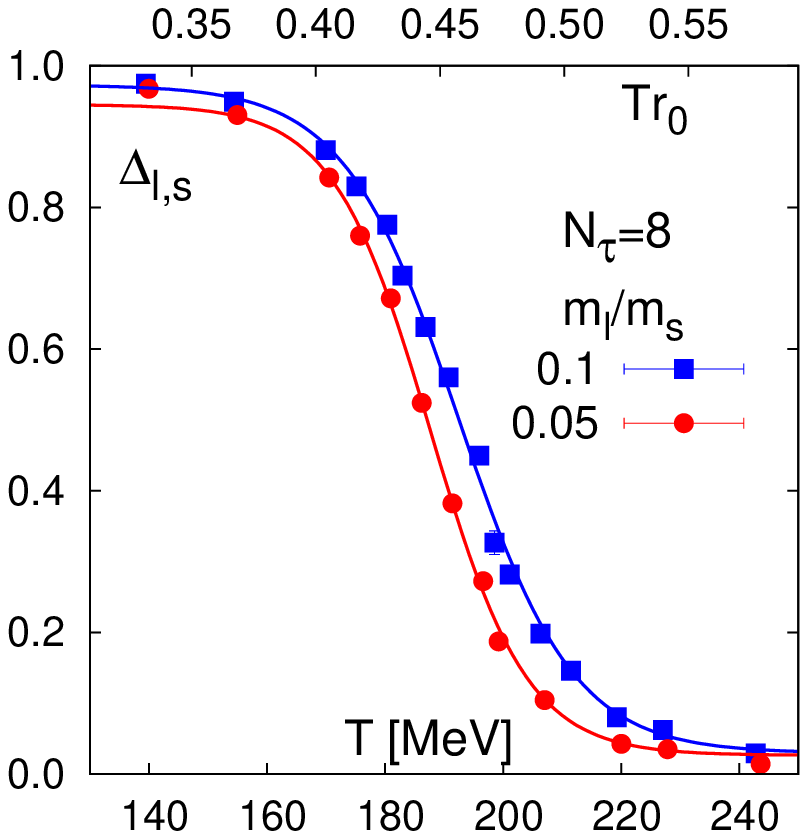}
\caption{The renormalized Polyakov loop (top) and the subtracted chiral condensate (bottom) as function of the
temperature calculated at $m_l=0.05m_s$ and at $0.1m_s$.}
\label{fig:orderpar}
\end{figure}

In Figure \ref{fig:orderpar} 
we show the renormalized Polyakov loop
and the subtracted chiral condensate $\Delta_{l,s}$ and compare with previous calculations performed at light
quark masses equal to one tenth of the strange quark mass \cite{hotqcd_eos}.
The renormalized Polyakov loop rises in the temperature interval $T=(170-200)$ MeV
where we also see the rapid increase of
the energy density. At the same time the subtracted chiral condensate rapidly drops in the 
transition region, indicating
that the approximate restoration of the chiral symmetry happens in the same temperature interval as deconfinement.
Compared to the calculation performed at light quark masses equal to one tenth of the strange quark mass we see
a shift of the transition region by roughly $5$ MeV. 
We note that such a shift arises differently in different observables.
In the case of the subtracted chiral condensate, for instance, a major ingredient to the
'shift' is the fact, that at fixed temperature
the condensate in the transition region is strongly quark mass dependent and drops
proportional to $\sqrt{m_l/m_s}$ \cite{Goldstone}.

\begin{figure}
\includegraphics[width=7cm]{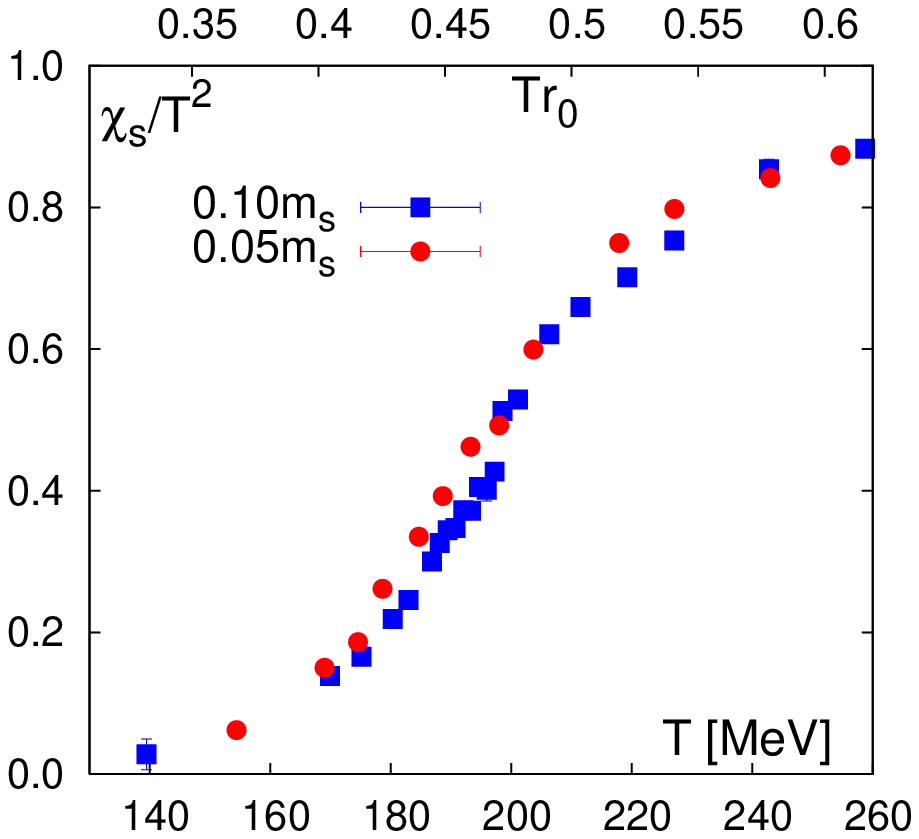}
\includegraphics[width=7cm]{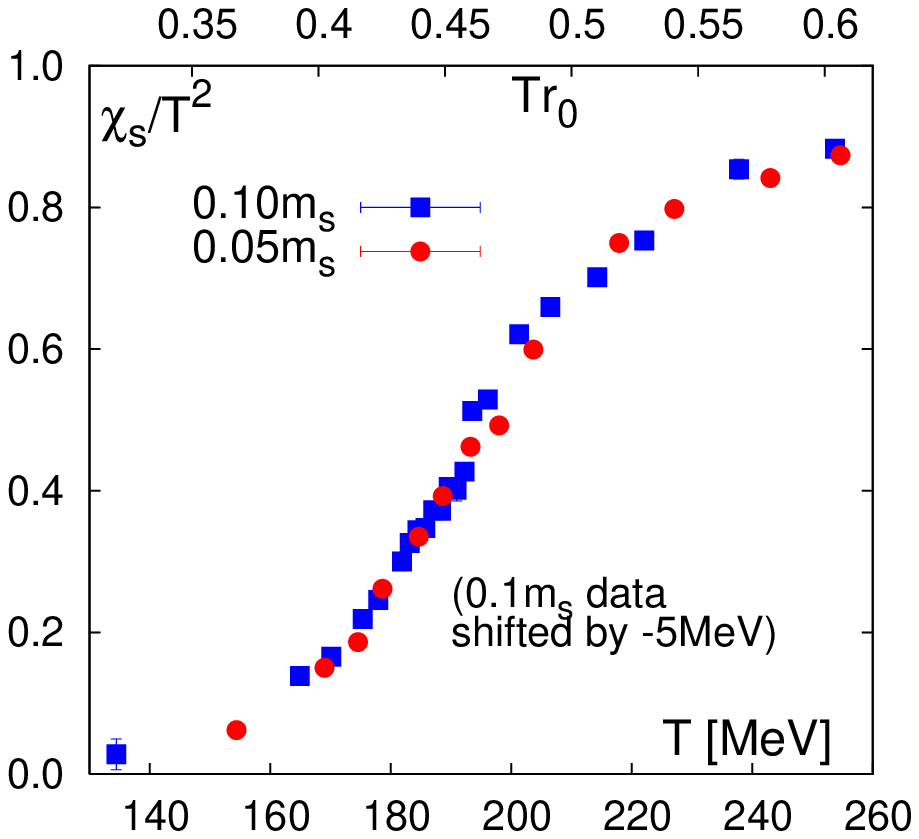}
\caption{Strangeness fluctuations as function of the
temperature calculated at $m_l=0.05m_s$ and at $0.1m_s$.
In the bottom figure the numerical data for $m_l=0.1m_s$ have been shifted
by $5$ MeV.
}
\label{fig:chis}
\end{figure}

The fluctuation of strangeness is also indicative of deconfinement. It can be defined
as the second derivative of the free energy density with respect to the strange quark
chemical potential
\begin{equation}
\chi_s(T)=\frac{1}{T^3 V}\frac{\partial^2 \ln Z(T,\mu_s)}{\partial \mu_s^2}|_{\mu_s=0}.
\end{equation}
At low temperatures 
strangeness is carried by massive hadrons and therefore strangeness fluctuations are
suppressed. At high temperatures strangeness is carried by quarks and 
the effect of the strange quark mass is small. Therefore strangeness fluctuations are not suppressed
at high temperatures. As discussed in Ref. \cite{hotqcd_eos} strangeness fluctuations behave
like the energy density in the transition region, i.e. they rapidly rise in a narrow temperature 
interval. In Fig. \ref{fig:chis} we show the strangeness fluctuations calculated at $m_l=0.05m_s$
and compare them with previous calculations performed at $m_l=0.1m_s$ \cite{hotqcd_eos}.
In the  bottom figure we also show the strangeness fluctuation for $m_l=0.1m_s$ with a
$5$ MeV shift of the temperature scale. As one can see this shift accounts for most of the 
quark mass dependence of the strangeness fluctuations. This is consistent with the conclusion
obtained from the quark mass dependence of other thermodynamic observables. 

\section{Conclusion}

We have calculated the EoS, renormalized Polyakov loop, subtracted chiral condensate and
strangeness fluctuations in (2+1)-flavor QCD in the crossover region from low to
high temperatures using the
improved p4 staggered fermion formulation on lattices with temporal extent $N_{\tau}=8$
at physical values of the light and strange quark masses. 
We found that thermodynamic quantities below the deconfinement 
transition are larger compared to the previous
calculations performed at twice larger quark mass but fall below the
resonance gas model result. 
The differences in the thermodynamic quantities calculated at
$m_l=0.05m_s$ and $m_l=0.1m_s$ can be well understood in terms 
of the shift of the transition temperatures
towards smaller values when the quark mass is decreased.
This conclusion is also supported by the calculation of renormalized
Polyakov loop, subtracted chiral condensate and strangeness fluctuations.
No additional enhancement of the pressure and the energy density is 
seen at low temperatures. 
This and the deviation from the resonance gas model
may be a cutoff effect due to taste violations. However, better 
statistical accuracy and calculations at smaller lattice spacing 
are needed to quantify this assertion.
At temperatures above $200$ MeV no quark mass dependence 
is seen in the equation of state.

\section*{Acknowledgments}
\label{ackn}
This work has been supported in part by contracts DE-AC02-98CH10886
and DE-FG02-92ER40699 with the U.S. Department of Energy,
the Bundesministerium f\"ur Bildung und Forschung under grant
06BI401, the Gesellschaft
f\"ur Schwerionenforschung under grant BILAER, the Helmholtz Alliance
HA216/EMMI and the Deutsche
Forschungsgemeinschaft under grant GRK 881. Numerical simulations have
been performed
on the QCDOC computer of the RIKEN-BNL research center, the DOE funded
QCDOC at BNL, the apeNEXT at Bielefeld University and the BlueGene/L
at the New York Center for Computational Sciences (NYCCS). 

\section*{Appendix}
In this appendix we present some numerical details of our calculations.
In Table \ref{tab:bare} we give the expectation values of the gauge action and
quark condensates calculated on $32^4$ (zero temperature) and $32^3 \times 8$ 
(finite temperature) lattices. In Table \ref{tab:thermo} we present the numerical
values of the trace anomaly, pressure, energy density, bare Polyakov loop and strangeness
fluctuations. 
\begin{table*}
\begin{tabular}{|c|c|c|c|c|c|c|c|c|c|}
\hline
$\beta$ & $\hat m_s$ & $T$ [MeV] &  \# traj. & $\langle s_G \rangle_0$ &  $\langle s_G \rangle_{\tau}$ &
$\langle\bar \psi \psi  \rangle_{l,0}$ & $\langle\bar \psi \psi  \rangle_{l,\tau}$ &
$\langle\bar \psi \psi  \rangle_{s,0}$ & $\langle\bar \psi \psi  \rangle_{s,\tau}$  \\
\hline
3.4300 & 0.0370 & 139 &  7950 & 4.10978(19) & 4.10922(15) & 0.06996(19) & 0.06788(12) & 0.14253(12) & 0.1419(8) \\
3.4600 & 0.0313 & 154 & 15290 & 4.04326(16) & 4.04242(17) & 0.05263(14) & 0.04932(9)  & 0.11670(8)  & 0.1157(8) \\
3.4900 & 0.0290 & 170 & 33710 & 3.98357(19) & 3.98215(11) & 0.04018(13) & 0.03453(11) & 0.10023(10) & 0.0984(7) \\
3.5000 & 0.0253 & 175 & 32520 & 3.96333(14) & 3.96161(12) & 0.03538(12) & 0.02783(18) & 0.08873(9)  & 0.0862(9) \\
3.5100 & 0.0260 & 180 & 30050 & 3.94573(19) & 3.94322(7)  & 0.03294(13) & 0.02339(20) & 0.08733(10) & 0.0839(7) \\
3.5200 & 0.0240 & 185 & 39990 & 3.92713(9)  & 3.92431(10) & 0.02956(6)  & 0.01718(17) & 0.08003(5)  & 0.0756(8) \\
3.5300 & 0.0240 & 191 & 70230 & 3.90994(11) & 3.90651(6)  & 0.02715(9)  & 0.01250(16) & 0.07735(7)  & 0.0721(7) \\
3.5400 & 0.0240 & 196 & 56740 & 3.89333(10) & 3.88947(6)  & 0.02518(13) & 0.00929(8)  & 0.07507(8)  & 0.0691(5) \\
3.5450 & 0.0215 & 198 & 20700 & 3.88427(6)  & 3.88048(9)  & 0.02328(6)  & 0.00681(12) & 0.06848(3)  & 0.0618(9) \\
3.5600 & 0.0205 & 206 & 11660 & 3.85944(14) & 3.85579(7)  & 0.02039(11) & 0.00459(6)  & 0.06315(8)  & 0.0559(8) \\
3.5850 & 0.0192 & 219 & 17250 & 3.81987(14) & 3.81650(10) & 0.01638(8)  & 0.00297(2)  & 0.05583(7)  & 0.0479(8) \\
3.6000 & 0.0192 & 227 &  4260 & 3.79720(7)  & 3.79408(17) & 0.01478(7)  & 0.00272(2)  & 0.05359(4)  & 0.0460(9) \\
3.6300 & 0.0170 & 243 & 10810 & 3.75307(9)  & 3.75040(9)  & 0.01141(8)  & 0.00202(7)  & 0.04524(6)  & 0.0378(4) \\
3.6600 & 0.0170 & 259 & 45050 & 3.71067(8)  & 3.70856(3)  & 0.00916(11) & 0.00184(2)  & 0.04202(5)  & 0.0355(2) \\
\hline
\end{tabular}

\caption{The expectation values of the gauge action, light and strange quark condensates
at zero and finite temperatures. Also shown are the temperature values for each value of gauge
coupling $\beta$ obtained from $r_0=0.469$fm. We also give the number of MD trajectories for each run}
\label{tab:bare}
\end{table*}
\begin{table*}
\begin{tabular}{|c|c|c|c|c|c|}
\hline
$T$ [MeV] & $(\epsilon-3 p)/T^4$ &  $p/T^4$ &  $\epsilon/T^4$ & $L_{bare}$ &	$\chi_s/T^2$ \\
\hline	
139  &  0.89(30) &  0.067 &   1.09(30)  &   0.00087(6) &  -             \\
154  &  1.34(30) &  0.161 &   1.82(30)  &   0.0015(5)  &  0.062(4)      \\
170  &  2.34(30) &  0.317 &   3.29(30)  &   0.0028(6)  &  0.150(9)      \\
175  &  2.87(27) &  0.393 &   4.05(27)  &   0.0036(9)  &  0.186(10)     \\
180  &  4.23(29) &  0.495 &   5.71(29)  &   0.0046(9)  &  0.261(10)     \\
185  &  4.87(19) &  0.621 &   6.73(39)  &   0.0057(6)  &  0.335(6)      \\
191  &  5.99(19) &  0.768 &   8.30(19)  &   0.0069(7)  &  0.392(10)     \\
196  &  6.84(18) &  0.929 &   9.63(18)  &   0.0082(4)  &  0.462(5)      \\
198  &  6.78(18) &  1.013 &   9.81(18)  &   0.0089(9)  &  0.492(7)      \\
206  &  6.73(27) &  1.274 &  10.55(27)  &   0.0104(11) &  0.599(11)     \\
219  &  6.48(31) &  1.703 &  11.59(31)  &   0.0135(17) &  0.750(10)     \\
227  &  6.15(34) &  1.926 &  11.93(34)  &   0.0146(24) &  0.798(8)      \\
243  &  5.49(26) &  2.296 &  12.38(26)  &   0.0172(11) &  0.841(7)      \\
259  &  4.58(18) &  2.602 &  12.38(18)  &   0.0199(5)  &  0.874(3)      \\
\hline
\end{tabular}

\caption{The numerical values of the trace anomaly, pressure, energy density, bare Polyakov loop and strangeness
fluctuations at each temperature.}
\label{tab:thermo}
\end{table*}


\vspace*{-0.6cm}

\begin{thebibliography}{99}

\bibitem{su2}
J.~Engels, F.~Karsch, H.~Satz and I.~Montvay,
  Phys.\ Lett.\  B {\bf 101}, 89 (1981) and 
Nucl.\ Phys.\  B {\bf 205}, 545 (1982).
\bibitem{su3}
J.~B.~Kogut, H.~Matsuoka, M.~Stone, H.~W.~Wyld, S.~H.~Shenker, J.~Shigemitsu and 
D.~K.~Sinclair,
  Phys.\ Rev.\ Lett.\  {\bf 51}, 869 (1983);\\
T.~Celik, J.~Engels and H.~Satz,
  Phys.\ Lett.\  B {\bf 129}, 323 (1983).


\bibitem{detar_lat08}
  C.~E.~DeTar,
  PoS {\bf LAT2008}, 001 (2008).

\bibitem{petreczky_sewm06}
  P.~Petreczky,
  Nucl.\ Phys.\ Proc.\ Suppl.\  {\bf 140}, 78 (2005).


\bibitem{petreczky_qm09}
  P.~Petreczky,
  Nucl.\ Phys.\  A {\bf 830}, 11C (2009).

\bibitem{stout}
  Y.~Aoki, Z.~Fodor, S.~D.~Katz and K.~K.~Szabo,
  JHEP {\bf 0601}, 089 (2006).

\bibitem{MILCeos}
C.~Bernard {\it et al.},
Phys.\ Rev.\  D {\bf 75}, 094505 (2007).

\bibitem{ourEoS}
M.~Cheng {\it et al.},
Phys.\ Rev.\  D {\bf 77}, 014511 (2008).

\bibitem{hotqcd_eos}
A.~Bazavov {\it et al.},
Phys.\ Rev.\  D {\bf 80}, 014504 (2009).

\bibitem{FlavBreak}
M.~Cheng {\it et al.}, Eur. Phys. J. C {\bf 51}, 875 (2007).

\bibitem{ourTc}
  M.~Cheng {\it et al.},
  Phys.\ Rev.\  D {\bf 74}, 054507 (2006).


\bibitem{gray}
A. Gray {\it et al.}, Phys. Rev. D {\bf 72}, 094507 (2005).

\bibitem{allton}
C. Allton, Nucl. Phys. B [Proc. Suppl.] {\bf 53}, 867 (1997).


\bibitem{milc01}
  C.~W.~Bernard {\it et al.},
  Phys.\ Rev.\  D {\bf 64}, 054506 (2001).

\bibitem{milc04}
  C.~Aubin {\it et al.},
  Phys.\ Rev.\  D {\bf 70}, 094505 (2004).


\bibitem{cleymans}
J.~Cleymans and K.~Redlich,
Phys.\ Rev.\  C {\bf 60}, 054908 (1999).

\bibitem{pbm}
A.~Andronic, P.~Braun-Munzinger and J.~Stachel,
Nucl.\ Phys.\  A {\bf 772}, 167 (2006).


\bibitem{fodorTc}
 Y.~Aoki, Z.~Fodor, S.~D.~Katz and K.~K.~Szabo,
  Phys.\ Lett.\  B {\bf 643}, 46 (2006);
  Y.~Aoki, S.~Borsanyi, S.~Durr, Z.~Fodor, S.~D.~Katz, S.~Krieg and K.~K.~Szabo,
  JHEP {\bf 0906}, 088 (2009).

\bibitem{stickan}
  F.~Karsch, C.~Schmidt and S.~Stickan,
  Comput.\ Phys.\ Commun.\  {\bf 147}, 451 (2002).

\bibitem{mclerran81}
L.D.McLerran and B.Svetitsky, Phys. Rev. D {\bf 24}, 450 (1981).

\bibitem{okacz02}
  O.~Kaczmarek, F.~Karsch, P.~Petreczky and F.~Zantow,
  Phys.\ Lett.\  B {\bf 543}, 41 (2002)

\bibitem{plqcd}
  P.~Petreczky and K.~Petrov,
  Phys.\ Rev.\  D {\bf 70}, 054503 (2004);
  O.~Kaczmarek and F.~Zantow,
  Phys.\ Rev.\  D {\bf 71}, 114510 (2005).

\bibitem{Goldstone}
  S.~Ejiri {\it et al.},
  Phys.\ Rev.\  D {\bf 80}, 094505 (2009).



\end{thebibliography}
\end{document}